\RequirePackage[mathlines]{lineno} 
\documentclass[a4paper,11pt]{article}
\pdfoutput=1 

\usepackage{jheppub} 
\usepackage{overpic}
\usepackage{amssymb}
\usepackage{amsmath}
\usepackage[T1]{fontenc} 

\usepackage{makecell}
\usepackage{graphicx}
\usepackage{subfigure}
\usepackage{epstopdf}

\usepackage{threeparttable}
\usepackage{multirow}
\usepackage{subfigure}
\usepackage{float}
\usepackage{LamcNPi-defs}
\let\oldequation\equation
\let\oldendequation\endequation
\renewenvironment{equation}
  {\linenomathNonumbers\oldequation}
  {\oldendequation\endlinenomath}

\begin{document}

\title{\bf \boldmath
First observation of $\LamCstarA \to \LamC\pi^0\pi^0$ and $\LamCstarB\to \LamC\pi^0\pi^0$}
\collaborationImg{\includegraphics[height=4cm,angle=90 ]{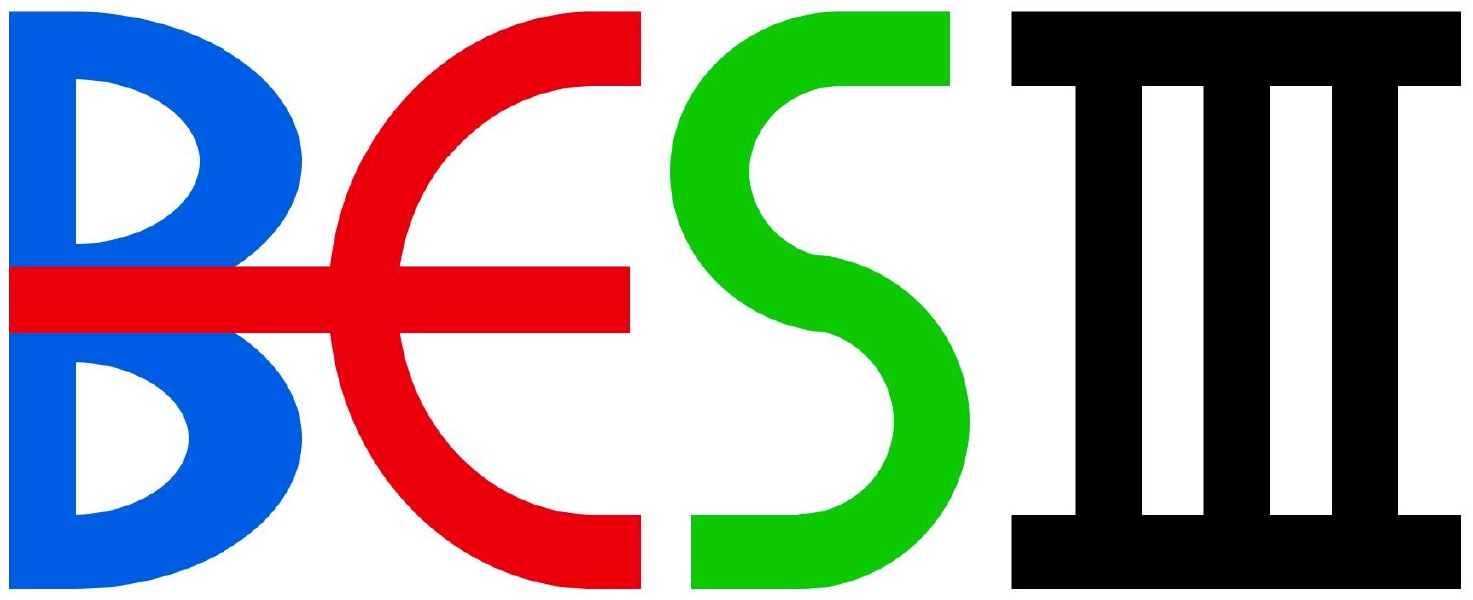}}

\collaboration{BESIII Collaboration}

\author{
M.~Ablikim$^{1}$, M.~N.~Achasov$^{4,c}$, P.~Adlarson$^{76}$, O.~Afedulidis$^{3}$, X.~C.~Ai$^{81}$, R.~Aliberti$^{35}$, A.~Amoroso$^{75A,75C}$, Y.~Bai$^{57}$, O.~Bakina$^{36}$, I.~Balossino$^{29A}$
% ...
\\
\vspace{0.2cm}
(BESIII Collaboration)\\
\vspace{0.2cm} {\it
$^{1}$ Institute of High Energy Physics, Beijing 100049, People's Republic of China\\
% ...
$^{3}$ Bochum  Ruhr-University, D-44780 Bochum, Germany\\
$^{4}$ Budker Institute of Nuclear Physics SB RAS (BINP), Novosibirsk 630090, Russia\\
% ...
$^{29}$ INFN Sezione di Ferrara, (A)INFN Sezione di Ferrara, I-44122, Ferrara, Italy; (B)University of Ferrara,  I-44122, Ferrara, Italy\\
% ...
$^{35}$ Johannes Gutenberg University of Mainz, Johann-Joachim-Becher-Weg 45, D-55099 Mainz, Germany\\
$^{36}$ Joint Institute for Nuclear Research, 141980 Dubna, Moscow region, Russia\\
% ...
$^{57}$ Southeast University, Nanjing 211100, People's Republic of China\\
% ...
$^{75}$ University of Turin and INFN, (A)University of Turin, I-10125, Turin, Italy; (B)University of Eastern Piedmont, I-15121, Alessandria, Italy; (C)INFN, I-10125, Turin, Italy\\
$^{76}$ Uppsala University, Box 516, SE-75120 Uppsala, Sweden\\
% ...
$^{81}$ Zhengzhou University, Zhengzhou 450001, People's Republic of China\\

\vspace{0.2cm}
% ...
$^{c}$ Also at the Novosibirsk State University, Novosibirsk, 630090, Russia\\
% ...

}
}

\abstract{
By analysing $e^+e^-$ annihilation data corresponding to an integrated luminosity of 368.48 \ipb{}
collected at the centre-of-mass energies of $\sqrt{s} = 4.918$ and $4.951$~GeV with the BESIII detector, 
we report the first observation of $\LamCstarA$ and $\LamCstarB\to \LamC\pi^0\pi^0$ with statistical significances of 7.9$\sigma$ and 11.8$\sigma$, respectively.  
The branching fractions of $\LamCstarA$ and $\LamCstarB\to \LamC\pi^0\pi^0$ are
measured to be $\sbf$ and $\dbf$, respectively. The absolute branching fraction of $\LamCstarA$ is consistent with the expectation of the mechanism referred to as the \textit{threshold effect}, proposed for the strong decays of $\LamCstarA$ within uncertainty.
}

\maketitle
\flushbottom
\section{INTRODUCTION}

Charmed baryon spectroscopy provides an ideal platform for studying the dynamics of light quarks in the presence of a heavy quark.
The strong decays of charmed baryons are most conveniently described by heavy hadron chiral perturbation theory,
in which heavy quark symmetry and chiral symmetry are incorporated~\cite{PhysRevD.75.014006,PhysRevD.92.074014}.
The chiral Lagrangian involves several coupling constants for transitions between $s$-wave and $p$-wave 
charmed baryons, referred to as $h_2 - h_{15}$ defined in Ref.~\cite{PhysRevD.46.1148,PhysRevD.56.5483}. 
Among these, 
the coupling constants $h_2$ and $h_8$ can be extracted from the strong decays of $\LamCstarA$ and $\LamCstarB$~\cite{Cheng:2015Front,Cheng:2021qpd}.
These couplings are critical to describe the charmed baryon spectrum and to make predictions of decays into other charmed baryons. However, to date, the strong decays of $\LamCstarA$ and $\LamCstarB$ remain poorly 
understood due to the lack of experimental data~\cite{PDG:2022}. The existing determinations of $h_2$ and $h_8$ 
are based on the measured decay widths of $\LamCstarA$ and $\LamCstarB$.
Since the width of $\LamCstarB$ is nearly zero~\cite{PDG:2022,PhysRevD.107.032008}, 
only the upper limit on $h_8$ is provided.
A precise measurement of the strong decays of $\LamCstarA$ and $\LamCstarB$ is highly desirable. The corresponding branching fractions (BFs) are important inputs for determining $h_2$ and $h_8$.

In the quark model, the $\LamCstarA$ and $\LamCstarB$ are the lowest-lying excited states of $\LamC$ 
with spin-parities of $1/2^{-}$ and $3/2^{-}$, respectively, and form a degenerate pair of the 
$p$-wave state~\cite{Cheng:2015Front,Cheng:2021qpd}.
The upper limit on the absolute BF of the decay $\LamCstarA\to \LamC\pi^+\pi^-$ 
was determined to be less than $85.0\%$~\cite{changjie} at the $90\%$ confidence level.
The absolute BF of the process $\Lambda_{c}(2625)^{+}\to\LamC\pi^+\pi^-$ has been measured to be 
$(51.1 \pm 5.8_{\rm stat.} \pm 3.5_{\rm syst.}) \%$~\cite{changjie}.
However, the absolute BFs of these $\pi^0\pi^0$ transitions have, until now, never been measured experimentally. 
Given that only an upper limit on the absolute BF has been obtained for the decay $\LamCstarPiPiAp$, 
measuring the absolute BF of the decay $\LamCstarA\to \LamC\pi^0\pi^0$ becomes increasingly important.
Assuming isospin symmetry, the ratio between the BFs of $\pi^+\pi^-$ and $\pi^0\pi^0$ transitions is 2:1, 
which forms the basis for the BFs of various strong $\LamCstarA$ and $\LamCstarB$ decays quoted in the 
Particle Data Group (PDG)~\cite{PDG:2022}.
However, isospin symmetry in these processes has not been verified by any experimental measurement. 
In Ref.~\cite{PhysRevD.67.074033}, a mechanism known as the \textit{threshold effect} is proposed to account for 
the limited transition phase space in these strong decays. If this mechanism applies here as well, 
it would break the 2:1 relation between $\pi^+\pi^-$ and $\pi^0\pi^0$ transitions in $\LamCstarA$ decay.


In this paper, we report the first measurement of the absolute BFs of $\LamCstarA$ and 
$\LamCstarB\to\LamC\pi^0\pi^0$, obtained from the processes of 
$\ee\to\ALamC\LamCstarA+c.c.$ and $\ALamC\LamCstarB+c.c.$.
We use the data collected with the BESIII detector at centre-of-mass (c.m.) energies 
$\sqrt s = 4.918$ and 4.951~GeV~\cite{Baician:2023}. 
The integrated luminosities of the data samples at 4.918 and 4.951~GeV are 
208.1 and 160.4~pb$^{-1}$~\cite{BESIII:Lumi}, respectively.

\section{BESIII DETECTOR AND MONTE CARLO SIMULATION}

The BESIII detector~\cite{Ablikim:2009aa} records symmetric $e^+e^-$ collisions
provided by the BEPCII~\cite{BEPCII} storage ring,
which operates in the center-of-mass energy ($\sqrt{s}$) range from 1.84 to
4.95~GeV, with a peak luminosity of $1 \times
10^{33}\;\text{cm}^{-2}\text{s}^{-1}$ achieved at $\sqrt{s} =
3.77\;\text{GeV}$.
BESIII has collected large data samples at these energy regions~\cite{detector1}.
The cylindrical core of the BESIII detector covers 93\% of the full$~$ solid angle and
consists of a helium-based multilayer drift chamber~(MDC), a plastic scintillator time-of-flight
system~(TOF), and a CsI(Tl) electromagnetic calorimeter~(EMC),
which are all enclosed in a superconducting solenoidal magnet providing a
{\spaceskip=0.2em\relax 1.0 T} magnetic field~\cite{Kaixuan:2022}. The solenoid is supported by an
octagonal flux-return yoke with resistive plate counter based muon
identification modules interleaved with steel.
The charged-particle momentum resolution at $1~{\rm GeV}/c$ is $0.5\%$,
and resolution of the ionization energy loss in the MDC ($\mathrm{d}E/\mathrm{d}x$)
is $6\%$ for electrons
from Bhabha scattering. The EMC measures photon energies with a
resolution of $2.5\%$ ($5\%$) at $1$~GeV in the barrel (end-cap) region.
The time resolution in the TOF barrel region is
68 ps, while that in the end-cap region is 110 ps.
The end-cap TOF system was upgraded in 2015 using multi-gap
resistive plate chamber technology, providing a time
resolution of 60 ps~\cite{detector2}.

Simulated samples are produced with a Geant4-based~\cite{Agostinelli:2002hh} Monte Carlo (MC) toolkit,
which includes a full implementation of the detector geometry and response~\cite{Kaixuan:2022} of the BESIII detector. The simulations are used to determine the efficiency of the detector and the reconstruction, and to estimate the background.
The inclusive MC samples, which consists of \LCLC{} events, $D_{(s)}$ production,
$\psi$ states produced in initial state radiation processes, and continuum processes $\ee\to q\bar{q}$
($q=u,d,s$), is generated to estimate the potential background. Here, all the known decay modes of charmed hadrons and charmonia are modeled with
{\sc evtgen}~\cite{Lange:2001uf, Ping:2008zz} using BFs taken from the PDG~\cite{PDG:2022}, while the remaining unknown decays are modeled with
{\sc lundcharm}~\cite{Chen:2000tv,PhysRevLett.31.061301}.
Final state radiation from charged final state particles is incorporated using
{\sc photos}~\cite{Richter-Was:1992hxq}.

\section{MEASUREMENT METHOD}

This analysis is performed to search for the signal processes 
$\Lambda_{c}(2595)^{+}\to \Lambda^{+}_{c} \pi^{0} \pi^{0}$
and $\Lambda_{c}(2625)^{+}\to \Lambda^{+}_{c} \pi^{0} \pi^{0}$,
based on the productions $\ee\to\ALamC\LamCstarA$ and $\ALamC\LamCstarB$, 
and to measure their BFs in a model-independent approach.
Firstly, three hadronic decay modes are used to reconstruct the $\LamCb$ candidates, 
which are referred to as tagged $\LamCb$ hereafter.
These three hadronic decay modes are $\LamCb \to \pkpib$, $\pksb$, and $\bar{\Lambda}\pi^-$, 
where the subsequent decay modes of the intermediate states are $\Ks\to\pi^+\pi^-$ and 
$\bar{\Lambda}\to \bar{p}\pi^+$. Then, we search for the signals 
$\LamCstar \to \Lambda^{+}_{c} \pi^{0} \pi^{0}$ on the recoiling side against the $\LamCb$. 
Here, $\LamCstar$ represents either $\Lambda_{c}(2595)^{+}$ or $\Lambda_{c}(2625)^{+}$.

The BFs of $\LamCstar \to \LamC\pi^0\pi^0$ are determined as,
\begin{equation} \label{eq:method}
  \mathcal{B}=\frac{(N_{\LamCstar}^{\mathrm{obs}} - N_{\rm bkg}) \cdot \sum_{i} \mathcal{B}_{\mathrm{tag}}^{i} \epsilon_{\mathrm{tag}}^{i} } { N_{\mathrm{tag}} \cdot \sum_{i}\mathcal{B}_{\mathrm{tag}}^{i} \epsilon_{\mathrm{sig}}^{i} },
\end{equation}
where $i$ represents each tag mode, $\mathcal{B}_{\mathrm{tag}}$ denotes the BFs of the tag modes,
$\epsilon_{\mathrm{tag}}$ corresponds to the efficiencies of reconstructing the tagged  $\LamCb$, and 
$\epsilon_{\mathrm{sig}}$ represents the efficiencies of selecting the signals 
$\LamCstar \to \Lambda^{+}_{c} \pi^{0} \pi^{0}$ together with the tagged $\LamCb$,
as listed in Table~\ref{tab:results} and Table~\ref{tab:results1}. 
The $N_{\mathrm{tag}}$ stands for the number of total events for $\LamCstarA$ or $\LamCstarB$ after 
selecting the tagged $\LamCb$. 
The $N_{\LamCstar}^{\mathrm{obs}}$ is the number of observed events for signal 
$\LamCstar \to \Lambda^{+}_{c} \pi^{0} \pi^{0}$.
Due to the fact that the energy of the photons is always lower than 150~MeV, the efficiency of detecting all 
four $\gamma$'s from the two $\pi^0$'s is only 10\%. 
Therefore, we have adopted a partial reconstruction strategy to increase the efficiency of selecting the signal
$\LamCstar \to \Lambda^{+}_{c} \pi^{0} \pi^{0}$, rather than full reconstruction. 
The peaking background contamination originated from the sources $\LamCstar \to \LamC \pi^+\pi^-$ 
will be subtracted, and the corresponding number of events is denoted as $N_{\mathrm{bkg}}$.

The signal MC samples of $\ee\to\ALamC\LamCstarA / \ALamC\LamCstarB \to \ALamC \LamC\pi^0\pi^0$ 
are generated for the individual c.m.~energies using the generator {\sc kkmc}~\cite{Jadach:2000ir},
incorporating initial state radiation effects and beam energy spread.
To achieve a more accurate simulation, the polar angle ($\theta$) distributions of $\LamCstarA$ and $\LamCstarB$ 
are considered in the generator via a parametrization of $f(\cos\theta) \propto (1+\alpha_{\Lambda_{c}}\cos^2\theta)$.
For $\ee\to\ALamC\LamCstarB$, the $\alpha_{\Lambda_{c}}$ is assigned to the measured values in
Ref.~\cite{Junhua:2023}, 
while for $\ee\to\ALamC\LamCstarA$,  the $\alpha_{\Lambda_{c}}$ is set to 1. 
Additionally,  any possible deviations from these values are considered as a source of systematic uncertainty.
The $\LamCb$ is required to decay into the three tag modes, while the $\LamC$ decays into any allowed final states.
The input line shapes for their cross sections are also obtained from Ref.~\cite{Junhua:2023}.
The charge-conjugate processes $\ee\to\LamC\ALamCstarA$ and $\LamC\ALamCstarB$ are similarly generated 
with $\ALamCstarA$ and  $\ALamCstarB \to \ALamC\pi^0\pi^0$,
where the $\LamC$ subsequently decays to any allowed processes and the $\LamCb$ decays to the three tag modes. These signal MC samples are used to extract the signal shapes and efficiencies 
$\epsilon_{\mathrm{sig}}^{i}$. 

Moreover, to extract the total yields $N_{\mathrm{tag}}$ and corresponding efficiencies $\epsilon_{\mathrm{tag}}^{i}$, 
additional MC samples of $\ee\to\ALamC\LamCstarA$ and $\ALamC\LamCstarB$ 
are generated with the $\LamCb$ decaying into the three tag modes and the $\LamCstar$ into any allowed processes.
Their charge-conjugate processes $\ee\to\LamC\ALamCstarA$ and $\LamC\ALamCstarB$ are produced 
with the $\LamC$ decaying into any allowed processes and the $\LamCb$ into the three tag modes. 
The decays of $\LamCstarA$ and $\LamCstarB$ are modelled based on the information in the PDG~\cite{PDG:2022},
with both decaying into $\LamC\pi^+\pi^-$ and $\LamC\pi^0\pi^0$ final states. 

In the analysis, the events with $\LamCb \to \pkpi$, $\pks$, and $\lampi$ are selected in the meantime, 
which correspond to the charge-conjugate modes of tagged $\LamCb$ and 
are referred to as tagged $\LamC$. 
Signal MC samples are also generated for the tagged $\LamC$, which are exactly the same as those for 
tagged $\LamCb$. For simplicity, throughout this paper, only cases with tagged $\LamCb$ are described,
while those with charge-conjugate mode of tagged $\LamC$ are implicitly included.

\section{SELECTION FOR TAGGED $\LamCb$ AND EXTRACTION OF THE NUMBER OF TOTAL EVENTS FOR $\LamCstar$} \label{sec:single-tag}
Charged tracks detected in the helium-based MDC are required to be within a polar angle ($\theta$) range of $\left|\rm{\cos}\theta\right|<0.93$, where $\theta$ is defined with respect to the $z$-axis. The distance of closest approach for charged tracks that do not come from a $\Lambda$ or $K_{S}^{0}$ decay are required to be within $\pm$10 cm along the $z$-axis and 1 cm in the plane perpendicular to the beam.
The particle identification (PID) is implemented by combining measurements of the specific ionization energy loss in the MDC and the TOF between the interaction point and the dedicated TOF detector system. Each charged track that doesn't come from $K_{S}^{0}$ and $\bar{\Lambda}$ is assigned a particle type of pion, kaon or proton, according to which assignment has the highest probability.
For the mode $\LamCb\to \bar{p}K^{+}\pi^{-}$, a vertex fit is performed to each $\bar{p} K^+\pi^-$ combination candidate, and the re-fitted momenta are used in the further study.

Candidates for $K_{S}^{0}$ and $\bar{\Lambda}$ are reconstructed by their dominant modes $K_{S}^{0} \to \pi^{+}\pi^{-}$ and $\bar{\Lambda} \to \bar{p}\pi^{+}$, respectively,
where the charged tracks are required to have distances of closest approaches to the interaction point that are within $\pm$20 cm along the $z$-axis.
To improve the signal purity, the PID requirement is only applied to the (anti)proton candidate, but not for the charged pion. A secondary vertex fit constrained by the decay vertex and the production vertex is performed to each $K_{S}^{0}$ or $\bar{\Lambda}$ candidate, and the re-fitted momenta are used in the further analysis.
The $K_{S}^{0}$ or $\bar{\Lambda}$ candidate is accepted by requiring the $\chi^2$ of the
secondary vertex fit to be less than 100. Furthermore, the
decay vertex is required to be separated from the interaction point by a
distance of at least twice the fitted vertex resolution, and the
invariant mass to be within (0.487, 0.511) GeV/$c^2$ for $\pi^{+}\pi^{-}$ and (1.111, 1.121) GeV/$c^2$ for $\bar{p}\pi^{+}$.

\begin{figure}[!htp]
    \begin{center}
    \vspace{-0.3cm}
    \subfigtopskip=0.5pt
    \subfigbottomskip=0.5pt
            \subfigure{\includegraphics[width=0.46\textwidth,height=0.24\textheight, trim=5 0 0 0, clip]{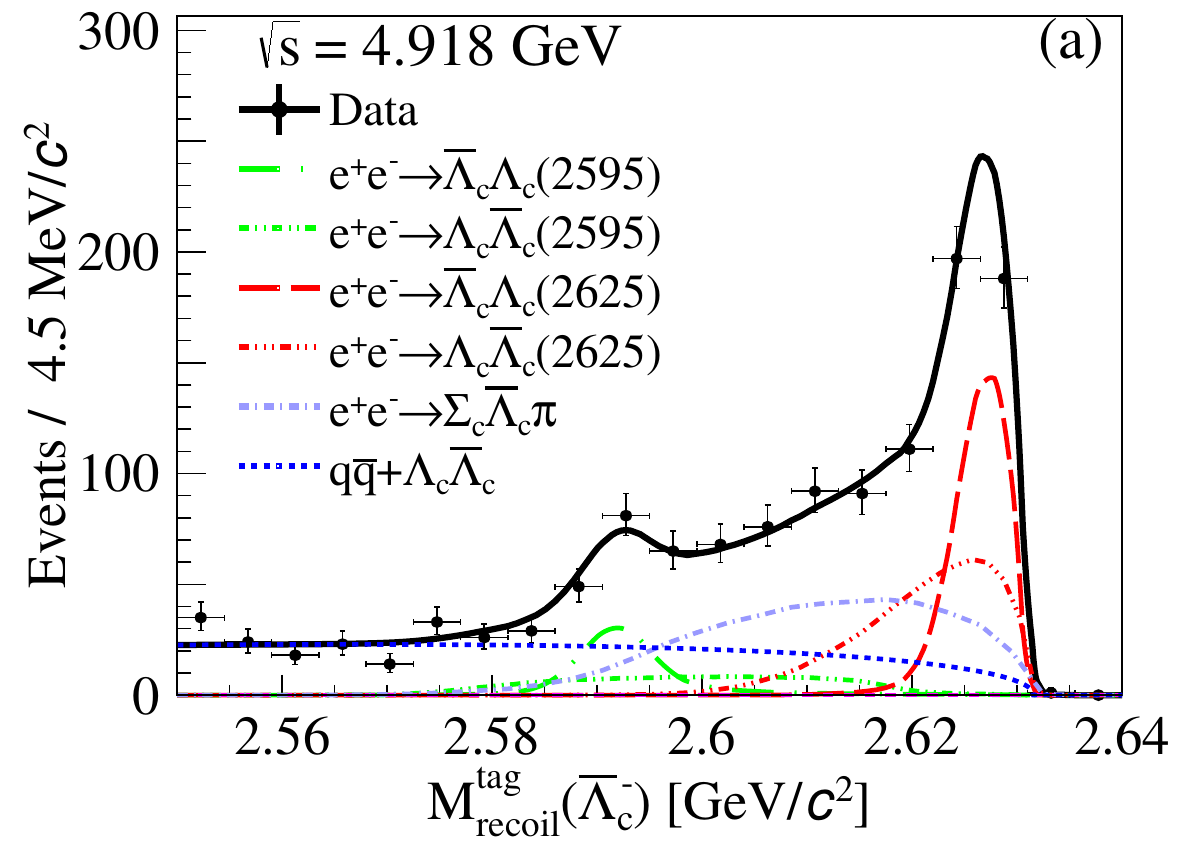}~\label{fig:tag1}} \hspace{-0.34cm}
            \subfigure{\includegraphics[width=0.46\textwidth,height=0.24\textheight, trim=5 0 0 0, clip]{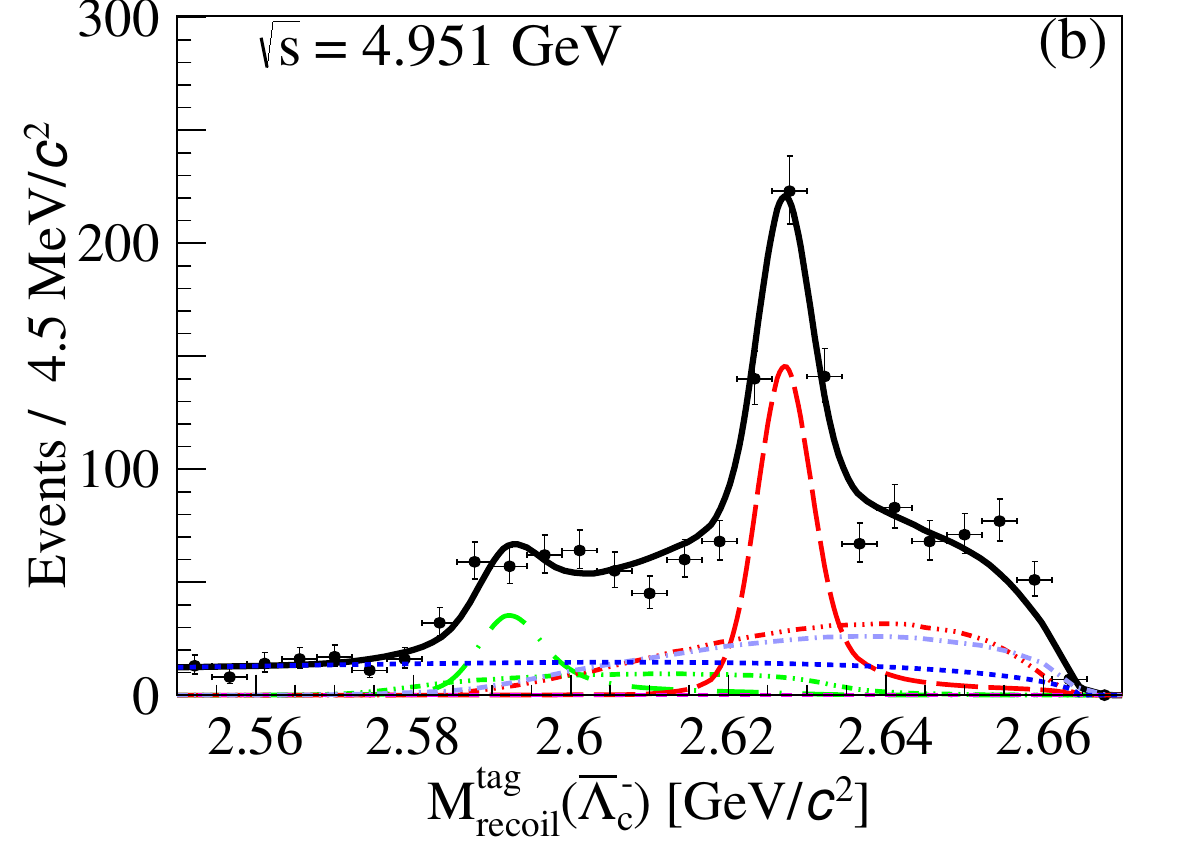}~\label{fig:tag2}}
	\end{center}
 \vspace{-0.6cm}
    \caption{
    The fits to the distributions $\MtagrecLc$ at $\sqrt s = 4.918$~GeV (a) and 4.951~GeV (b). The black points with error bars are data. The solid curves represent the fit results,
    and the dashed curves describe individual components including both signals and backgrounds.
    The contribution of $e^+e^- \to \Sigma_{c}\bar{\Sigma}_{c}$ is negligible according to the fit results and
    cannot be seen evidently in the figures.
       }
        \label{fig:tag-fit}
\end{figure}

All combinations for each tag mode of $\LamCb$ are retained, and their invariant mass, $M(\LamCb)$, 
is required to fall within the range (2.27, 2.30)~GeV/$c^2$. The recoiling mass of the tagged $\LamCb$,  
$\MtagrecLc$, is depicted in \figurename~\ref{fig:tag1} and \ref{fig:tag2} by combining the three tag modes. 
The resonances $\LamCstarA$ and $\LamCstarB$ are observed at both energy points, indicating the existence of processes $\ee\to\LamCb\LamCstarA$ and $\LamCb\LamCstarB$.
Meanwhile, the charge-conjugate processes $\ee\to\LamC\ALamCstarA$ and $\LamC\ALamCstarB$ distribute broadly
beneath the resonances, as the tagged $\LamCb$ can originate from the decays of $\ALamCstarA$ or $\ALamCstarB$. 
These productions, where the tagged $\LamCb$ arises directly from $\ee$ collision, are denoted as 
$S_{ee}^{\rm tag}$, while those where the the tagged $\LamCb$ originates from the decays of 
$\bar{\Lambda}_{c}^{\ast -}$ are denoted as $S_{\rm inte}^{\rm tag}$. 


An unbinned maximum likelihood fit is performed to the distribution of $\MtagrecLc$ (denoted as $\rm fit_{\rm tag}$
hereafter). Both processes $S_{ee}^{\rm tag}$ and $S_{\rm inte}^{\rm tag}$ are included, 
and they are modelled by MC simulations in the $\rm fit_{\rm tag}$.
The two contributions are correlated with the same production cross section and individual detection efficiencies 
$\epsilon_{\mathrm{tag}}$.
The contributions $S_{ee}^{\rm tag}$ are further convolved with Gaussian functions, which are shared between the two resonances due to the limited sample sizes at individual energy values, 
and account for the resolution difference between data and MC simulation.
The backgrounds include contributions from continuum hadron production (denoted as $q\bar{q}$), 
$\ee \to\Lambda^+_c\bar{\Lambda}^-_c$, $\ee \to \Sigma_{c}\bar{\Sigma}_{c}$, 
and $\ee \to \Sigma_{c}\ALamC\pi$,
where $\pi$ denotes the isospin triplets $\pi^\pm$ and $\pi^0$,
and the $\Sigma_{c}$ denotes the triplets $\Sigma_{c}^{++}$, $\Sigma_{c}^{+}$, and $\Sigma_{c}^{0}$. 
The processes $q\bar{q}$ and $\Lambda^+_c\bar{\Lambda}^-_c$ are grouped together and 
modelled by ARGUS functions~\cite{ARGUS:1990}, the shape parameters of which are obtained 
by fitting the distributions of $\MtagrecLc$ in the data at $\sqrt s=4.840$~GeV~\cite{BESIII:Lumi},
which is below the $\LamCstar$ production threshold.
The magnitudes of the ARGUS functions are free parameters in the $\rm fit_{\rm tag}$.
The shapes of the contributions $\ee \to \Sigma_{c}\bar{\Sigma}_{c}$ and $\Sigma_{c}\ALamC\pi$ are derived 
from MC simulations, and their yields are determined in the $\rm fit_{\rm tag}$. The fit results are depicted in \figurename~\ref{fig:tag1} and \ref{fig:tag2}, where the background contributions of 
$\ee \to \Sigma_{c}\bar{\Sigma}_{c}$ are negligible. 
The significances of  $\LamCstarA$ components are 5.3$\sigma$ and 8.3$\sigma$ for 
$\sqrt{s} = 4.918$ and 4.951~GeV, respectively. The corresponding values for $\LamCstarB$ components 
are 12.7$\sigma$ and 14.0$\sigma$.

\begin{table}[htbp]\centering
\caption{
 The detection efficiencies, $\epsilon_{\rm tag}$ and $\epsilon_{\rm sig}$, for each tag mode at 
 $\sqrt{s}=4.918$~GeV.
  The numbers of events, $N_{\mathrm{tag}}$ and $N_{\mathrm{sig}}$ 
  ($N_{\LamCstar}^{\mathrm{obs}} - N_{\rm bkg}$), combine the three tag modes.}\label{tab:results}
  \renewcommand\arraystretch{1.2}\footnotesize
\begin{tabular}{|c|c|c|c|c|c|}
\hline
      & $\LamCb$ decays & $\epsilon_{\mathrm{tag}}$ (\%) & $\epsilon_{\mathrm{sig}}$ (\%) & $N_{\mathrm{tag}}$ & $N_{\mathrm{sig}}$ \\\cline{1-6}
    \multirow{3}{2 cm}{$\ALamC\LamCstarA$} & $\pkpi$                  & 48.5 & 27.2 & &\\ \cline{2-4}
    & $\pks$                   & 49.9 & 26.7 & 135.2 $\pm$ 26.4 & 40.3 $\pm$ 7.6\\ \cline{2-4}
    & $\bar{\Lambda}\pi^-$     & 38.5 & 20.7 & & \\
\hline
\multirow{3}{2 cm}{$\ALamC\LamCstarB$} & $\pkpi$                  & 46.6 & 27.9 & &\\ \cline{2-4}
    & $\pks$                   & 50.0 & 26.7 & 418.7 $\pm$ 34.4 & 88.7 $\pm$ 11.3\\ \cline{2-4}
    & $\bar{\Lambda}\pi^-$     & 38.3 & 19.9 & & \\
    \hline
\end{tabular}
\end{table}

\begin{table}[htbp]\centering
\caption{
The detection efficiencies, $\epsilon_{\rm tag}$ and $\epsilon_{\rm sig}$, for each tag mode at $\sqrt{s}=4.951$~GeV.
  The numbers of events, $N_{\mathrm{tag}}$ and $N_{\mathrm{sig}}$ 
  ($N_{\LamCstar}^{\mathrm{obs}} - N_{\rm bkg}$),  combine the three tag modes.}\label{tab:results1}
  \renewcommand\arraystretch{1.2}\footnotesize
\begin{tabular}{|c|c|c|c|c|c|}
\hline
    & $\LamCb$ decays & $\epsilon_{\mathrm{tag}}$ (\%) & $\epsilon_{\mathrm{sig}}$ (\%) & $N_{\mathrm{tag}}$ & $N_{\mathrm{sig}}$ \\\cline{1-6}
   \multirow{3}{2 cm}{$\ALamC\LamCstarA$} & $\pkpi$                  & 48.8 & 22.3 & &\\ \cline{2-4}
    & $\pks$                   & 49.0 & 22.3 & 210.3 $\pm$ 28.3 & 54.5 $\pm$ 10.2\\ \cline{2-4}
    & $\bar{\Lambda}\pi^-$     & 37.8 & 16.6 & & \\
\hline
   \multirow{3}{2 cm}{$\ALamC\LamCstarB$} & $\pkpi$                  & 47.4 & 21.9 & &\\ \cline{2-4}
    & $\pks$                   & 49.6 & 22.4 & 670.9 $\pm$ 55.6 & 114.0 $\pm$ 14.6\\ \cline{2-4}
    & $\bar{\Lambda}\pi^-$     & 37.6 & 16.6 & & \\
\hline
\end{tabular}
\end{table}

\section{THE SELECTION OF $\Lambda^{+}_{c} \pi^{0} \pi^{0}$ CANDIDATES AND EXTRACTION OF THE SIGNAL YIELDS}

Since the energy of the photons from the two $\pi^0$'s
in the signal process $\LamCstar \to \LamC \pi^0\pi^0$ is less than 150~MeV, 
the efficiency of detecting all of them is only 10\%. 
To improve the efficiency, a partial reconstruction strategy, relying on the topological differences,  
is adopted to select the signals and suppress the dominant backgrounds.
Due to the presence of the tagged $\LamCb$, the dominant backgrounds to the signals
originate from the processes $\LamCstar \to \LamC \gamma$ and $\LamCstar \to \LamC \pi^+\pi^-$. 
Another potential background $\LamCstar \to \LamC \pi^0$ is not taken into account
because it is isospin violated process and its branching fraction is highly suppressed.

To suppress the background from $\LamCstar \to \Lambda^{+}_{c} \gamma$ decay,
we veto the events with $E_\gamma \in (0.28, 0.35)$~GeV for $\sqrt{s} = 4.918$~GeV
and $E_\gamma \in (0.25, 0.38)$~GeV for $\sqrt{s} = 4.951$~GeV, where $E_\gamma$ represents
the deposited energy of photons in the EMC.
To suppress the background from $\LamCstar \to \LamC \pi^{+} \pi^{-}$ decay,
we require the number of charged pions with $P_{\pi^{\pm}} < 150$~MeV/$c$ to be zero ($N_{\pi^{\pm}}=0$). 
Here, $P_{\pi^{\pm}}$ represents the momentum of $\pi^{+}$ or $\pi^{-}$ candidates, and
the identification criteria for $\pi^{\pm}$ are the same as those used in the reconstruction of the tagged $\LamCb$.
Additionally, we veto events if the number of $\pi^0$ with $P_{\pi^0} < 0.15$~GeV/$c$ is zero ($N_{\pi^0}>0$),  
where $P_{\pi^0}$ denotes the momentum of $\pi^{0}$. 
The $\pi^0$ is reconstructed with its dominant decay mode $\pi^0\to \gamma\gamma$,
and the photon candidates are identified using showers in the EMC. The deposited energy of each shower must be 
greater than 25~MeV in the barrel region ($|\cos\theta| \le 0.80$) or greater than 50~MeV in the end-cap region 
($0.86 \le |\cos\theta| \le 0.92$). To suppress electronic noise and showers unrelated to the event, 
the difference between the EMC time and the event start time is required to be within (0, 700)~ns. 
The $\pi^0$ candidates are formed with a photon pair within the invariant-mass region (0.115, 0.150)~GeV/$c^2$. 
To improve the resolution, a kinematic fit is performed by constraining the invariant mass of the photon pair to be the 
$\pi^0$ mass~\cite{PDG:2022} and requiring the corresponding $\chi^2$ of the fit to be less than 200.

\begin{figure}[!htp]
    \begin{center}
    \vspace{-0.3cm}
    \subfigtopskip=0.5pt
    \subfigbottomskip=0.5pt
            \subfigure{\includegraphics[width=0.46\textwidth,height=0.24\textheight, trim=5 0 0 0, clip]{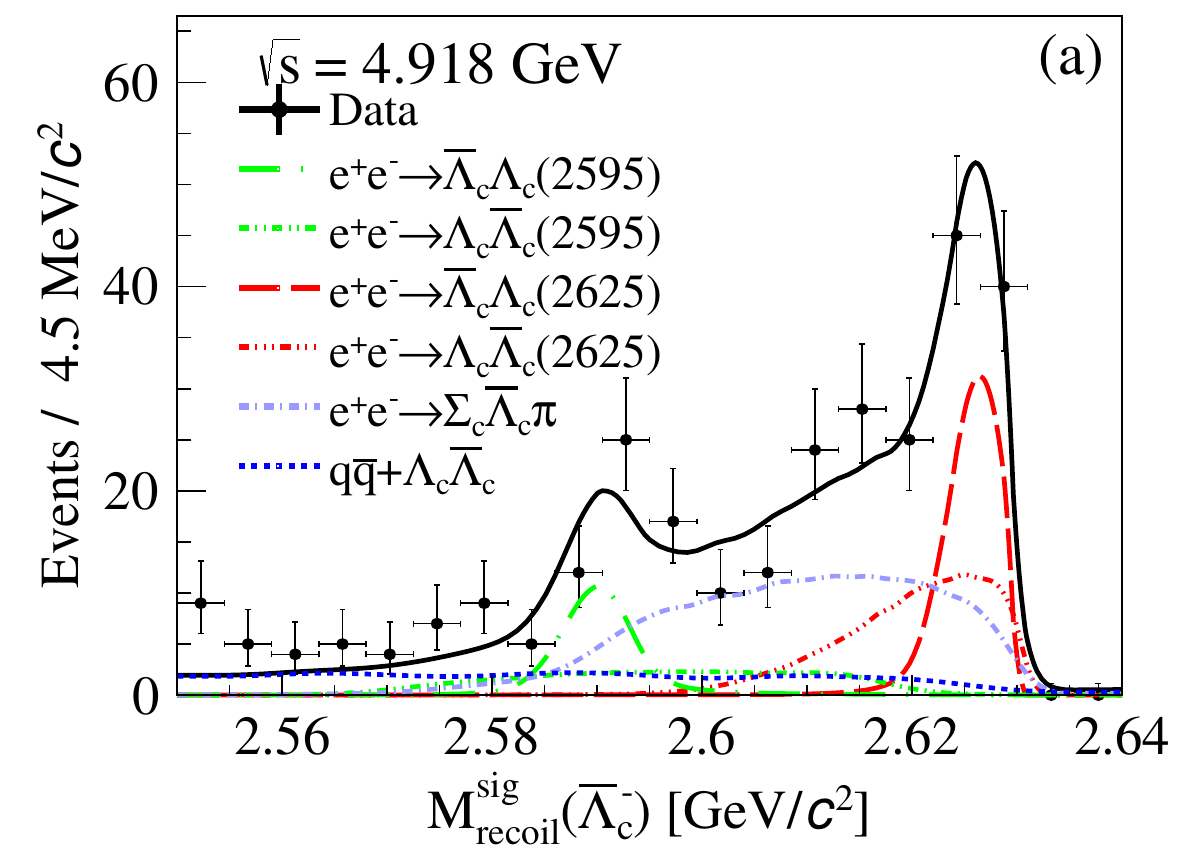}~\label{fig:sig1}} \hspace{-0.34cm}
            \subfigure{\includegraphics[width=0.46\textwidth,height=0.24\textheight, trim=5 0 0 0, clip]{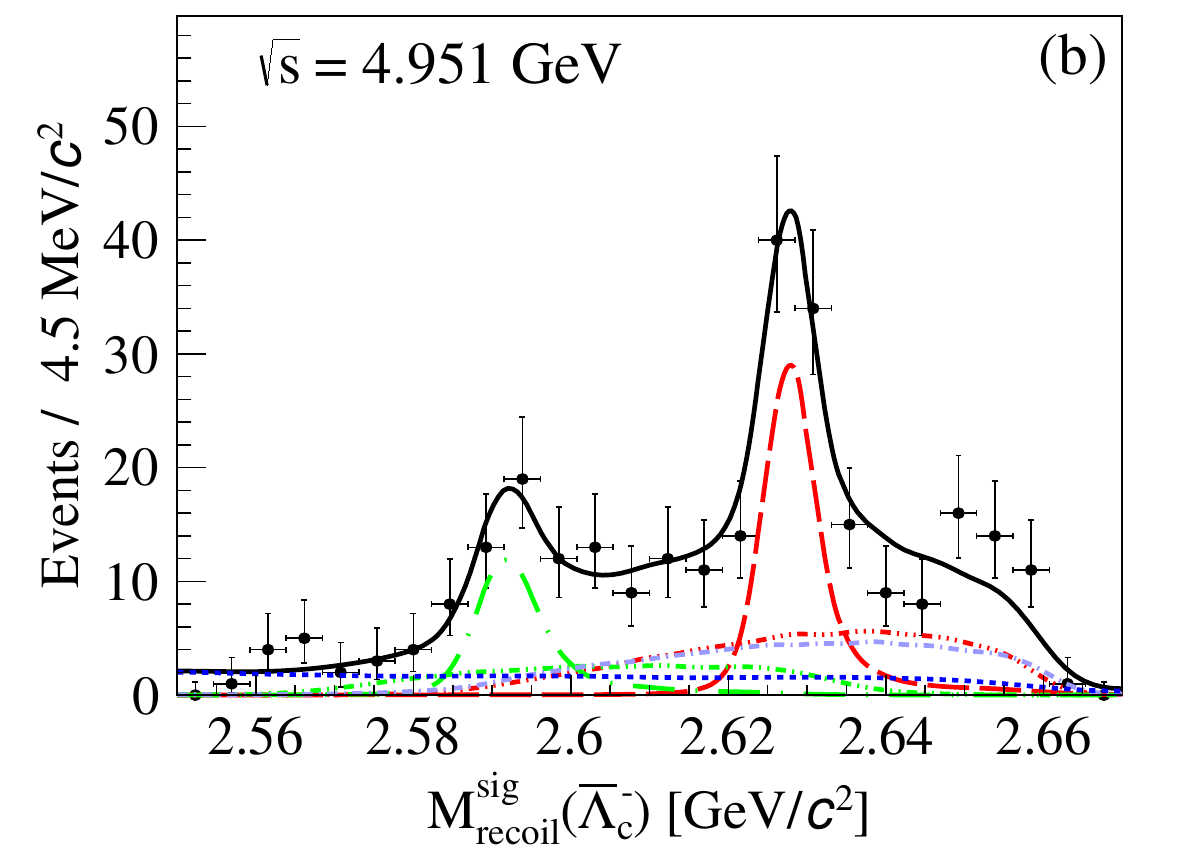}~\label{fig:sig2}}
	\end{center}
 \vspace{-0.6cm}
    \caption{
    The fits to the distributions $M^{\rm sig}_{\rm recoil}(\LamCb)$ at $\sqrt s = 4.918$~GeV (a) and 4.951~GeV (b). The black points with error bars are data. The solid curves represent the fit results,
    and the dashed curves describe individual components including both signals and backgrounds.
       }
        \label{fig:sig-fit}
\end{figure}

With all the above selections, the signals $\LamCstar \to \LamC \pi^0\pi^0$ are further investigated 
in the recoiling mass distribution against the tagged $\LamCb$, $M^{\rm sig}_{\rm recoil}(\LamCb)$, 
as shown in \figurename~\ref{fig:sig-fit} for both energy points.
Here, the resonances $\LamCstarA$ and  $\LamCstarB$ are clearly observed, 
indicating the signals $\LamCstarA$ and  $\LamCstarB\to \LamC \pi^0\pi^0$. 
Similarly to \figurename~\ref{fig:tag-fit}, the signal processes with the tagged $\LamCb$ arising directly from 
$\ee$ collision are denoted as $S_{ee}^{\rm sig}$, while those with the tagged $\LamCb$ from the  
decays of $\bar{\Lambda}_{c}^{\ast -}$ are denoted as $S_{\rm inte}^{\rm sig}$. 
Another unbinned maximum likelihood fit is performed to the distribution of $M^{\rm sig}_{\rm recoil}(\LamCb)$ (denoted as $\rm fit_{\rm sig}$ hereafter), where both signal contributions 
$S_{ee}^{\rm sig}$ and $S_{\rm inte}^{\rm sig}$ are included, and their shapes are modelled by 
MC simulations. 
The sources of non-peaking backgrounds are the same as those in the previous $\rm fit_{\rm tag}$. 
The processes $q\bar{q}$ and $\Lambda^+_c\bar{\Lambda}^-_c$ are 
modelled using inclusive MC samples in the $\rm fit_{\rm sig}$, with their magnitudes determined 
by fitting the events in the sideband region of $M(\LamCb)\in [2.19,2.25]$~GeV/$c^{2}$ and $M(\LamCb)\in [2.32,2.38]$~GeV/$c^{2}$. 
The shapes of the $e^+e^- \to \Sigma_{c}\ALamC\pi$ backgrounds are modelled by MC simulations,
with their magnitudes treated as free parameters in the $\rm fit_{\rm sig}$. 
Here, the process $e^+e^- \to \Sigma_{c}\bar{\Sigma}_{c}$ is not included, as its contribution is already negligible in the previous ${\rm fit}_{\rm tag}$. 
The resulting fitted curves are depicted in \figurename~\ref{fig:sig1} and \ref{fig:sig2}, and
the statistical significances of  signals $\LamCstarA$ and $\LamCstarB \to \LamC \pi^0\pi^0$ are 5.3$\sigma$ and 8.3$\sigma$ 
by combining the two energy points.


Since the $\LamCstar \to\Lambda^{+}_{c}\pi^+\pi^-$ backgrounds exhibit similar distributions in 
$M^{\rm sig}_{\rm recoil}(\bar{\Lambda}^-_c)$ as the signals $\LamCstar \to\Lambda^{+}_{c}\pi^0\pi^0$, 
they also contribute events to the resonances $\LamCstarA$ and  $\LamCstarB$. 
These backgrounds will be subtracted from the signal yields in the $\rm fit_{\rm sig}$, denoted as the term ``$N_{\LamCstar}^{\mathrm{obs}} - N_{\rm bkg}$'' in Equation~\ref{eq:method}.
The $\LamCstar \to\Lambda^{+}_{c}\pi^+\pi^-$ backgrounds are estimated using the corresponding 
exclusive MC samples, and their event yields are normalised with 
$\br(\LamCstarPiPiAp) = (66.0 \pm 14.1_{\rm stat.}) \%$~\cite{PDG:2022} and 
$\br(\LamCstarPiPiBp) = (51.1 \pm 5.8_{\rm stat.} \pm 3.5_{\rm syst.}) \%$~\cite{changjie}, respectively.
This results into 7.5 and 5.7 events for $\Lambda_{c}(2595)^+\to\Lambda^{+}_{c}\pi^+\pi^-$ background,
and 5.5 and 3.4 events for $\Lambda_{c}(2625)^+\to\Lambda^{+}_{c}\pi^+\pi^-$ background at 
$\sqrt{s} = 4.918$~GeV and $\sqrt{s} = 4.951$~GeV, respectively. 
Another potential source of peaking backgrounds is the processes $\LamCstar \to\Lambda^{+}_{c}\gamma$.
However, the decay $\LamCstar \to\Lambda^{+}_{c}\gamma$ has not been observed~\cite{PDG:2022}.
Therefore, the contributions are estimated by conservatively assigning
$\br(\Lambda_{c}(2595)^+\to\Lambda^{+}_{c}\gamma) = 5.0 \%$ and 
$\br(\Lambda_{c}(2625)^+\to\Lambda^{+}_{c}\gamma) = 5.0 \%$. This results into 1.7 and 1.3 events for 
$\Lambda_{c}(2595)^+\to\Lambda^{+}_{c}\gamma$, 
and 1.3 and 0.8 for $\Lambda_{c}(2625)^+\to\Lambda^{+}_{c}\gamma$ at $\sqrt{s} = 4.918$ and $4.951$~GeV, respectively.
The contamination from $\LamCstar \to\Lambda^{+}_{c}\gamma$ backgrounds is considered negligible.
The obtained signal yields are listed in Table~\ref{tab:results}.

Due to inconsistencies in the information of fake photons between data and MC simulation, 
the efficiency of the requirement on $N_{\pi^0}$, obtained from signal MC samples, needs to be corrected. 
A control sample of $\eea \to \LCpair$ with $\LamCb$ decaying into the three tag modes and $\LamC$ to any allowed processes is selected using the data set at $\sqrt s= 4.68$~GeV. This control sample exhibits a photon environment 
comparable to that of the signal $\ee \to \LamCb \LamCstar$ with $\LamCstar \to \LamC\pi^0\pi^0$ and $\LamCb$ decaying into the three tag modes, excluding the $\gamma$'s from the signal $\pi^0$'s. 
To replicate the photon environment in data for the signal processes, a mixed sample is created. 
Each event in the mixed sample comprises non-signal $\gamma$'s from the control sample and 
signal $\gamma$'s from the decay of the $\pi^0$'s, which are the daughter particles of $\LamCstar$. 
The signal $\gamma$'s are obtained from the signal MC samples. 
All the non-signal and signal $\gamma$'s are randomly selected from the control sample and signal MC samples 
at the reconstruction level, respectively, and they are combined to form an event in the mixed sample. 
The $\pi^0$ candidates are then reconstructed with any photon pairs in each event of the mixed sample. 
Subsequently, the requirement on $N_{\pi^0}$ is applied to both the signal MC samples and the mixed sample, 
and the corresponding selection efficiencies are determined to be 0.2327 and 0.2132, respectively. Their ratio, 0.92, 
is assigned as the correction factor in the efficiencies $\epsilon_{\rm sig}$.

The BFs of $\Lcsgf$ and $\Lcsg$ are determined to be $\sbfs$ and $\dbfs$, respectively, where the uncertainties are statistical. Two toy MC samples are generated using the Probability Density Function (PDF) of $\br(\Lcsg)$ and $\br(\LamCstarPiPiBp)$ after cancelling the uncertainties from tag yields, and a set of the corresponding ratio of these two BFs is obtained. The ratio of the BFs between $\Lcsg$ and $\LamCstarPiPiBp$ is determined to be $0.8 \pm 0.2$.

\section{SYSTEMATIC UNCERTAINTY}
The systematic uncertainties in the BFs measurements primarily arise from the line shapes of cross sections, 
c.m.~energy, beam energy spread, tag yield, efficiency correction of the $N_{\pi^0}$ requirement, 
$\pi^\pm$ and $\gamma$ vetoes, background estimation, and fit strategy. According to the Equation~\ref{eq:method}, the selection criteria of the tagged $\LamCb$ affect both $\epsilon_{\rm tag}$ and $\epsilon_{\rm sig}$. Therefore, the systematic uncertainties in detection efficiency and $\mathcal{B}_{\mathrm{tag}}$ can be canceled out. 
All the systematic uncertainties are listed in Table~\ref{tab:systematic_all}.

\begin{itemize}
\item {\bf Line shapes of cross sections:}
The uncertainties associated with the input line shapes of cross sections for $e^+e^- \to \Lambda^+_c \bar{\Lambda}_c(2595)^-$ and $e^+e^- \to \Lambda^+_c \bar{\Lambda}_c(2625)^-$ are studied by generating alternative MC samples with the cross sections varied by $\pm$1$\sigma$~\cite{Junhua:2023} at $\sqrt{s}=4.918$ and 4.951~GeV. 
The two maximum differences in BFs with respect to the nominal values are 2.8\% and 2.1\% for 
$\Lambda_{c}(2595)^{+}$ and $\Lambda_{c}(2625)^{+}$, respectively.



\item {\bf Centre-of-mass energy:}
Since $\sqrt{s}=4.918$~GeV is close to the $\LamC\bar{\Lambda}_{c}(2625)^{-}$ threshold, the $f_{\rm ISR}$ factor 
is highly sensitive to its actual energy value, where $f_{\rm ISR}$ is initial state radiation correction. 
In this analysis, the c.m.~energy is measured to be $\sqrt{s}=4918.02\pm0.34\pm0.34$~MeV, and we obtain the systematic uncertainty in $f_{\rm ISR}$ at this point by varying the energy by $\pm 1\sigma$.  
The maximum difference in resultant efficiencies between this case and the nominal one is regarded as the systematic uncertainty. The uncertainties are 0.4\% and 3.3\% for $\Lambda_{c}(2595)^{+}$ and $\Lambda_{c}(2625)^{+}$, respectively. For the other energy value, which is far from the threshold, the systematic uncertainties are negligible. 
The uncertainties at two energy values are weighted by the cross sections~\cite{Junhua:2023} and are assigned as 
1.1\% and 0.1\% for $\Lambda_{c}(2595)^{+}$ and $\Lambda_{c}(2625)^{+}$, respectively.

\item {\bf Beam energy spread:}
The beam energy spread has been estimated to be 1.55 $\pm$ 0.18~MeV at $\sqrt{s}=4599.53$~MeV~\cite{jj}. 
Then, it is calculated to be 1.77 $\pm$ 0.20~MeV at $\sqrt{s}=4918.02$~MeV using the functions 
$\mathrm{BES}_{4918.02}=\frac{4918.02^{2}}{4599.53^{2}}\cdot \mathrm{BES}_{4599.53}$ and 
$\delta_{\mathrm{BES}_{4918.02}}=\frac{\delta_{\mathrm{BES}_{4599.53}}}{\mathrm{BES}_{4599.53}} \cdot \frac{4918.02^{2}}{4599.53^{2}}\cdot \mathrm{BES}_{4599.53}$ as described in Ref.~\cite{be}. 
We obtain the systematic uncertainty at this energy point by generating alternative MC samples 
with the the beam energy spread set to 1.57~MeV and 1.97~MeV. The difference in the resultant efficiencies is 
regarded as the systematic uncertainty. The uncertainties are 0.0\% and 3.4\% for $\Lambda_{c}(2595)^{+}$ and 
$\Lambda_{c}(2625)^{+}$, respectively. For the other energy point, the systematic uncertainties are negligible 
as it is far from the threshold. The uncertainties at the two energy points are weighted by the cross sections~\cite{Junhua:2023} and are assigned as 0.0\% and 1.1\% for $\Lambda_{c}(2595)^{+}$ and 
$\Lambda_{c}(2625)^{+}$, respectively.

\item {\bf Tag yields:}
The uncertainty due to the tag yields cancels out. However, the uncertainty arising from background fluctuations 
in the tag yields contributes an additional uncertainty, which is determined to be 
$\sqrt{\sigma_{N_{\rm tag}}-N_{\rm tag}}$, where $\sigma_{N_{\rm tag}}$ is the statistical uncertainty obtained from
the fit, and $N_{\rm tag}$ is the purely statistical uncertainty.
The uncertainties are 10.4\% and 5.2\% for $\Lambda_{c}(2595)^{+}$ and $\Lambda_{c}(2625)^{+}$, respectively.

 
\item {\bf Efficiency correction of the $N_{\pi^0}$ requirement:}
The uncertainty in the efficiency correction of the $N_{\pi^0}$ requirement arises from the statistical uncertainties 
in the yield of the mixed sample. 
The uncertainty, 1.2\%, is assigned as the systematic uncertainty.

\item {\bf $\pi^\pm$ and $\gamma$ vetoes:}
We study the uncertainty of $\pi^\pm$ and $\gamma$ vetoes together
using a control sample of $\bar{\Lambda}^{-}_{c} \Lambda^{+}_{c}$ pair at $\sqrt s=4.68$~GeV, 
where $\bar{\Lambda}^{-}_{c} \to \bar{p} K^{+} {\pi}^{-}$ and $\Lambda^{+}_{c}$ decays to any allowed processes.
After applying the $\pi^{\pm}$ and $\gamma$ vetoes, the efficiency difference of between data and MC simulation 
is 0.7\%, which is assigned as the systematic uncertainty.


\item {\bf Background estimation of $\LamCstar \to \LamC \pi^+\pi^-$:}
The uncertainty in background estimation of $\Lambda_{c}(2595)^+ \to \LamC \pi^+\pi^-$ is evaluated by estimating 
the event yields of $\Lambda_{c}(2595)^+\to\Lambda^{+}_{c}\pi^+\pi^-$ backgrounds using 
$\br(\Lambda_{c}(2595)^{+} \to \LamC\pi^+\pi^-) = (1-59.5\%)$. 
The new BF of $\Lambda_{c}(2595)^+\to\Lambda^{+}_{c}\pi^0\pi^0$ is measured to be 63.9\%. 
The difference of new BF and the nominal one is 7.4\%.
The uncertainty in background estimation of $\Lambda_{c}(2625)^+ \to \LamC \pi^+\pi^-$ is estimated by 
quoting the BF uncertainty of $\LamCstarPiPiBp = (50.7 \pm 5.0_{\rm stat.} \pm 4.9_{\rm syst.}) \%$ 
from Ref.~\cite{changjie}. The resultant uncertainties are 7.4\% and 1.0\% for $\Lambda_{c}(2595)^{+}$ and $\Lambda_{c}(2625)^{+}$, respectively.

\item {\bf Fit:}
     The uncertainty in the fitting strategy is also taken into account, which arises from:

     1) The uncertainty from the signal shape is estimated by changing the signal shape from the signal MC shape 
     to the signal MC shape convolved with Gaussian functions with free parameters. The difference, 0.5\%,
      is assigned as the uncertainty.

     2) The uncertainty arising from the estimation of the process $e^+e^-\to\Sigma_c\LamCBar\pi$:
We obtain the systematic uncertainty in the line shape of the $\Sigma_c \bar{\Lambda}_c \pi$ by 
generating alternative MC samples with the input line shapes of cross sections changed to a flat line shape. 
The difference in resultant BFs is regarded as the systematic uncertainty. The uncertainties are 0.02\% and 0.06\% 
for $\Lambda_{c}(2595)^{+}$ and $\Lambda_{c}(2625)^{+}$, respectively, and we consider these uncertainties negligible.
To estimate the uncertainty, we perform the fitting using only the shape of $\ee \to \Sigma _c^+\LamCBar\pi^0$ instead of the original shape of $e^+e^-\to\Sigma_c\LamCBar\pi$, and we fix the yields of 
$\ee\to\Sigma_c^+\bar{\Lambda}^{-}_c \pi^0$ based on the event number ratio of $\ee\to\Sigma_c$ and $\Sigma^+_c$. The ratio is obtained from the study of $\LamCstar \to \LamC \pi^+\pi^-$~\cite{changjie}. 
The deviations of the fitted signal yields are treated as the systematic uncertainties, 
which are 1.9\% and 5.6\% for $\LamCstarA$ and $\LamCstarB$, respectively.
\end{itemize}

Table~\ref{tab:systematic_all} summarises the sources of the systematic uncertainties
in the BF measurements of $\LamCstarA$ and $\LamCstarB\to \LamC\pi^0\pi^0$.
The total systematic uncertainties are calculated by adding all sources in quadrature, 
which are 13.3\% and 8.1\% for $\Lambda_{c}(2595)^{+}$ and $\Lambda_{c}(2625)^{+}$, respectively.

\begin{table}[htbp]
  \begin{center}
  \caption{Summary of systematic uncertainties (in \%).}\label{tab:systematic_all}
  \renewcommand\arraystretch{1.2}
    \begin{tabular}{ l c c}
      \hline
       Source &  \LamCstarPiPiA{} & \LamCstarPiPiB{} \\
	   \hline
       Line shapes of cross sections      & 2.8  & 2.1       \\
       C.m. energy           & 1.1  & 0.1       \\
       Beam energy spread    & 0.0  & 1.1       \\
       Tag yield              & 10.4 & 5.2 \\
       Efficiency correction of the $N_{\pi^0}$ requirement & 1.2 & 1.2\\
       $\pi^\pm$ and $\gamma$ vetoes & 0.7 & 0.7 \\
       Background estimation of $\LamCstar \to \LamC \pi^+\pi^-$ &  7.4 & 1.0             \\
        Fit               & 1.9 & 5.6 \\
       Total                     & 13.3 & 8.1 \\
      \hline
    \end{tabular}
  \end{center}
\end{table}

\section{SUMMARY}

The strong decays $\LamCstarPiPiA$ and $\LamCstarPiPiB$ are measured using the 368.48~$\ipb$ of 
$e^+e^-$ data collected at $\sqrt{s}=4.918$ and 4.951~GeV with the BESIII detector at BEPCII.
For the first time, the BF of $\LamCstarPiPiA$ is measured to be $\sbf$, 
while the BF of $\LamCstarPiPiB$ is measured to be $\dbf$.
The absolute BF of the process $\Lambda_{c}(2625)^{+}\to\LamC\pi^+\pi^-$ at $\sqrt{s}=4.918$ and 4.951 GeV 
has been measured to be $(51.1 \pm 5.8_{\rm stat.} \pm 3.5_{\rm syst.}) \%$~\cite{changjie}.
The upper limit on the BF of decay $\LamCstarPiPiAp$ is updated to be less than $58.0\%$ 
at the $90\%$ confidence level, based on the constraint 
$\br(\ensuremath{\Lambda_{c}(2595)^{+} \to \Lambda^{+}_{c} \pi^0\pi^0}) + \br(\ensuremath{\Lambda_{c}(2595)^{+} \to \Lambda^{+}_{c} \pi^+\pi^-})<=1$,
where the uncertainty of $\br(\ensuremath{\Lambda_{c}(2595)^{+} \to \Lambda^{+}_{c} \pi^0\pi^0})$ has been taken into account. 
Table~\ref{tab:results_summary} shows a comparison of our BFs with the results of the 
$\LamCstar \to \LamC \pi^+\pi^-$ decay.
This analysis provides a model-independent measurement, representing the first experimental results for 
$\LamCstarA$ and $\LamCstarPiPiB$. This results will serve as
an important inputs for calibrating relative measurements and guiding the search for unknown decays of 
$\Lambda_{c}(2595)^{+}$ and $\Lambda_{c}(2625)^{+}$.
The ratio of $\br(\LamCstarPiPiB)$ to $\br(\LamCstarPiPiBp)$ is calculated to be $0.8 \pm 0.2$ after cancelling the uncertainties from tag yields. 
The absolute BF of the process $\LamCstarPiPiA$ is consistent with the expected ``threshold effect" within the uncertainty.
However, further confirmation with additional data in the future is necessary.

\begin{table}[htbp]
  \begin{center}
  \caption{Summary of the results.}\label{tab:results_summary}
  \renewcommand\arraystretch{1.2}
  {
    \begin{tabular}{ l c c }
      \hline
          &  PDG~\cite{PDG:2022} & Measurements \\
      \hline
      $\br(\LamCstarPiPiA)$ &  $-$ &$\sbf$ \\
      \hline
       $\br(\LamCstarPiPiB)$ & $-$ &$\dbf$ \\
      \hline
      $\br(\LamCstarPiPiAp)$ &  $-$ &$< 85.0\%$~\cite{changjie} \\
      \hline
      $\br(\LamCstarPiPiAp)$ &  $-$ &$< 58.0\%$ estimated by this work \\
       \hline
       $\br(\LamCstarPiPiBp)$ & $\approx 67\%$ &  $(51.1\pm5.8_{\rm stat} \pm 3.5_{\rm syst})\%$~\cite{changjie} \\
       \hline
    \end{tabular}}
  \end{center}
\end{table}

\section{ACKNOWLEDGEMENT}
The BESIII Collaboration thanks the staff of BEPCII and the IHEP computing center for their strong support. This work is supported in part by National Key R\&D Program of China under Contracts Nos. 2023YFA1606000; 
National Natural Science Foundation of China (NSFC) under Contracts Nos. 11635010, 11735014, 11935015, 11935016, 11935018, 12025502, 12035009, 12035013, 12061131003, 12192260, 12192261, 12192262, 12192263, 12192264, 12192265, 12221005, 12225509, 12235017, 12361141819, 12475091; 
Guangzhou Navigation Project No. 2024A04J6334;
the Chinese Academy of Sciences (CAS) Large-Scale Scientific Facility Program; the CAS Center for Excellence in Particle Physics (CCEPP); Joint Large-Scale Scientific Facility Funds of the NSFC and CAS under Contract No. U1832207; 100 Talents Program of CAS; The Institute of Nuclear and Particle Physics (INPAC) and Shanghai Key Laboratory for Particle Physics and Cosmology; German Research Foundation DFG under Contracts Nos. FOR5327, GRK 2149; Istituto Nazionale di Fisica Nucleare, Italy; Knut and Alice Wallenberg Foundation under Contracts Nos. 2021.0174, 2021.0299; Ministry of Development of Turkey under Contract No. DPT2006K-120470; National Research Foundation of Korea under Contract No. NRF-2022R1A2C1092335; National Science and Technology fund of Mongolia; National Science Research and Innovation Fund (NSRF) via the Program Management Unit for Human Resources \& Institutional Development, Research and Innovation of Thailand under Contracts Nos. B16F640076, B50G670107; Polish National Science Centre under Contract No. 2019/35/O/ST2/02907; Swedish Research Council under Contract No. 2019.04595; The Swedish Foundation for International Cooperation in Research and Higher Education under Contract No. CH2018-7756; U. S. Department of Energy under Contract No. DE-FG02-05ER41374


\begin{thebibliography}{99}

\bibitem{PhysRevD.75.014006} H. Y. Cheng and C. K. C, \href{https://link.aps.org/doi/10.1103/PhysRevD.75.014006}{Phys. Rev. \textbf{D} \textbf{75}, 014006 (2007).}

\bibitem{PhysRevD.92.074014} H. Y. Cheng and C. K. C, \href{https://link.aps.org/doi/10.1103/PhysRevD.92.074014}{Phys. Rev. \textbf{D} \textbf{92}, 074014 (2015).}

\bibitem{PhysRevD.46.1148} T. M. Yan, H. Y. Cheng, C. Y. Cheung, G. L. Lin, Y. C. Lin, and H. L. Yu, \href{https://doi.org/10.1103/PhysRevD.46.1148}{Phys. Rev. \textbf{D} \textbf{46}, 1148 (1992).}

\bibitem{PhysRevD.56.5483} D. Pirjol and T. M. Yan, \href{https://doi.org/10.1103/PhysRevD.56.5483}{Phys. Rev. \textbf{D} \textbf{56}, 5483 (1997).}

\bibitem{Cheng:2015Front}
H. Y. Cheng,
\href{https://doi.org/10.1007/s11467-015-0483-z}{Front. Phys. {\bf 10}, 101406 (2015).}

\bibitem{Cheng:2021qpd}
H. Y. Cheng,
\href{https://doi.org/10.1016/j.cjph.2022.06.021}{Chin. J. Phys. {\bf 78}, 324 (2022).}

\bibitem{PDG:2022}
R. L. Workman {\it et al.} (Particle Data Group),
\href{https://doi.org/10.1093/ptep/ptac097}{Prog. Theor. Exp. Phys. {\bf 2022}, 083C01 (2022).}


\bibitem{PhysRevD.107.032008}
D. Wang \textit{et al.} (Belle Collaboration),
\href{https://link.aps.org/doi/10.1103/PhysRevD.107.032008}{Phys. Rev. \textbf{D} {\bf 107}, 032008 (2023).}

%
%
%

\bibitem{changjie}
M. Ablikim \textit{et al.} (BESIII Collaboration),
\href{https://arxiv.org/abs/2401.09225}{arXiv:2401.09225 (2024).}

\bibitem{PhysRevD.67.074033} E. B. Andrew and F. F. Adam and P. Dan and M. Y. John, \href{https://doi.org/10.1103/PhysRevD.67.074033}{Phys. Rev. \textbf{D} \textbf{67}, 074033 (2003).}



\bibitem{Baician:2023}
B.-C. Ke, J. Koponen, H. B. Li, and Y. H. Zheng,
\href{https://doi.org/10.1146/annurev-nucl-110222-044046}{ANNU. REV. NUCL. PART. S. {\bf 73}, 285 (2023).}

\bibitem{BESIII:Lumi} M. Ablikim \textit{et al.} (BESIII Collaboration), \href{https://dx.doi.org/10.1088/1674-1137/ac84cc}{Chin. Phys. \textbf{C} \textbf{46}, 113003 (2022).}

\bibitem{Ablikim:2009aa} M. Ablikim \textit{et al.} (BESIII Collaboration), \href{https://dx.doi.org/10.1016/j.nima.2009.12.050}{Nucl. Instrum. Meth. \textbf{A} \textbf{614}, 345 (2010).}

\bibitem{BEPCII}
C. H. Yu {\it et al.}, \href{https://doi.org/10.18429/JACoW-IPAC2016-TUYA01}{Proceedings of IPAC2016, Busan, Korea (JACoW, Geneva, Switzerland, 2016).}
\bibitem{detector1}
M. Ablikim {\it et al.} (BESIII Collaboration), \href{https://arxiv.org/pdf/1912.05983.pdf}{Chin. Phys. C {\bf 44}, 040001 (2020).}
\bibitem{Kaixuan:2022}
K. X. Huang \textit{et al.},
\href{https://doi.org/10.1007/s41365-022-01133-8}{Nucl. Sci. Tech. { \bf 33}, 142 (2022).}
\bibitem{detector2}
X. Li {\it et al.},
\href{https://link.springer.com/article/10.1007/s41605-017-0014-2}{Radiat. Detect. Technol. Methods {\bf 1}, 13 (2017);} Y. X. Guo {\it et al.}, \href{https://link.springer.com/article/10.1007/s41605-017-0014-2}{Radiat. Detect. Technol. Methods {\bf 1}, 15 (2017)}

\bibitem{Agostinelli:2002hh} S. Agostinelli \textit{et al.} (GEANT4 Collaboration), \href{https://dx.doi.org/10.1016/S0168-9002(03)01368-8}{Nucl. Instrum. Meth. \textbf{A} \textbf{506}, 250 (2003).}

\bibitem{Lange:2001uf} D. J. Lange, \href{https://dx.doi.org/10.1016/S0168-9002(01)00089-4}{Nucl. Instrum. Meth. \textbf{A} \textbf{462}, 152 (2001).}

\bibitem{Ping:2008zz} R. G. Ping, \href{https://dx.doi.org/10.1088/1674-1137/32/8/001}{Chin. Phys. \textbf{C} \textbf{32}, 599 (2008).}

\bibitem{Chen:2000tv} J. C. Chen, G. S. Huang, X. R. Qi, D. H. Zhang, and
Y. S. Zhu, \href{https://dx.doi.org/10.1103/PhysRevD.62.034003}{Phys. Rev. \textbf{D} \textbf{62}, 034003 (2000).}

\bibitem{PhysRevLett.31.061301}
Y. L. Yang, R. G. Ping, and H. Chen, \href{https://doi.org/10.1088/0256-307X/31/6/061301}{Phys. Rev. Lett. {\bf 31}, 061301 (2014).}

\bibitem{Richter-Was:1992hxq} E. R. Was, \href{https://dx.doi.org/10.1016/0370-2693(93)90062-M}{Phys. Lett. \textbf{B} \textbf{303}, 163 (1993).}

\bibitem{Jadach:2000ir} S. Jadach, B. F. L. Ward, and Z. Was, \href{https://dx.doi.org/10.1103/PhysRevD.63.113009}{Phys. Rev. \textbf{D} \textbf{63}, 113009 (2001).}


\bibitem{Junhua:2023} M. Ablikim \textit{et al.} (BESIII Collaboration), \href{https://journals.aps.org/prd/abstract/10.1103/PhysRevD.109.L071104}{Phys. Rev. \textbf{D} \textbf{109}, L071104 (2024).}

\bibitem{ARGUS:1990}  H. Albrecht \textit{et al.} (ARGUS Collaboration), \href{http://dx.doi.org/10.1016/0370-2693(90)91293-K}{Phys. Lett. \textbf{B} \textbf{241}, 278 (1990).}

\bibitem{jj} M. Ablikim \textit{et al.} (BESIII Collaboration), \href{https://inspirehep.net/literature?sort=mostrecent&size=25&page=1&q=PhysRevLett.120.132001}{Phys. Rev. Lett. {\bf 120}, 132001 (2018).}
\bibitem{be} M. Sands \textit{et al.},
    \href{https://www.slac.stanford.edu/pubs/slacreports/reports02/slac-r-121.pdf}{SLAC-r-121, (1970).}

%
\end{thebibliography}
\end{document}